\documentclass[12pt,preprint]{aastex}
\usepackage{amsmath}

\shorttitle{Radiative Transfer of Thermal Electron Synchrotron Polarization}

\shortauthors{Mao \& Wang}

\begin{document}


\title{Synchrotron Polarization Radiative Transfer: Relativistic Thermal Electrons Contribution}


\author{
Jirong Mao\altaffilmark{1,2,3}, Stefano Covino\altaffilmark{4}, and Jiancheng Wang\altaffilmark{1,2,3}
}
\altaffiltext{1}{Yunnan Observatories, Chinese Academy of Sciences, 650011 Kunming, Yunnan Province, China}
\altaffiltext{2}{Center for Astronomical Mega-Science, Chinese Academy of Sciences, 20A Datun Road, Chaoyang District, Beijing, 100012, China}
\altaffiltext{3}{Key Laboratory for the Structure and Evolution of Celestial Objects, Chinese Academy of Sciences, 650011 Kunming, China}
\altaffiltext{4}{Brera Astronomical Observatory, via Bianchi 46, I-23807, Merate (LC), Italy}

\email{jirongmao@mail.ynao.ac.cn}

\begin{abstract}
Relativistic thermal electrons moving in a large-scale magnetic field can produce synchrotron radiation. Linear synchrotron polarization can also be produced by the relativistic thermal electrons. In this paper, we utilize a hybrid thermal-nonthermal electron energy distribution to calculate circular synchrotron polarization. We further compute the radiative transfer of the synchrotron polarization in the optical and radio bands when we consider the contribution of the thermal electrons. We attempt to apply the polarization results to some astrophysical objects, such as kilonova like AT 2017gfo/GW170817, the fast radio burst (FRB), the Gamma-ray burst (GRB) afterglow, and the supernova remnant (SNR). The large optical depth of radiative transfer effects the small polarization degrees of these populations when the media surrounding the synchrotron sources take heavy absorption to the polarized photons. We need a strong magnetic field in our model to reproduce the linear and circular polarization properties that were observed in FRB 140514. This indicates that FRBs have a neutron star origin.
\end{abstract}




\section{Introduction}

Relativistic electrons moving in a large-scale and ordered magnetic field produce synchrotron radiation. If the relativistic electrons produced by the first-order Fermi acceleration have a nonthermal energy distribution, the synchrotron radiation turns out to be linearly polarized with a degree of $p_l=(3n+3)/(3n+7)$, where $n$ is the power-law index of the electron energy spectrum \citep{legg68,rybicki79}. This theory has often been applied to calculate the linear polarization in relativistic jets of several categories of celestial objects. In most cases, such as, e.g., for gamma-ray bursts (GRBs), the intrinsic polarization is affected by extrinsic effects, such as the magnetic field configuration, the jet structure, and the geometry between the ordered magnetic field and the jet direction. The resulting polarization can therefore be very different from that produced by the emission process alone (e.g., Granot \& K\"{o}nigl 2003; Nava et al. 2016).
When we consider the second-order Fermi acceleration, the relativistic electrons have a thermal Maxwellian component \citep{sch85,sch89a,sch89b,stawarz08,giannios09}. Recent numerical simulations have reproduced the thermal-nonthermal electron energy distribution in small length scales with the physics of magnetic reconnection and plasmoid instabilities \citep{cerutti14,nal15,sironi16,li17}. Therefore, when we consider a thermal electron component, the simple estimation of the synchrotron polarization degree is no longer valid.

\citet{jones79} calculated linear and circular synchrotron polarization of Maxwellian-distributed electrons. While the electron energy distribution is neither thermal nor nonthermal. A hybrid distribution of thermal-nonthermal electrons should be considered in reality. \citet{mao18} have computed the synchrotron polarization properties in detail when the hybrid thermal-nonthermal electron energy distribution is adopted. In that work, the linear polarization degrees dominated by the thermal/nonthermal electrons were presented in the radio, optical, X-ray and $\gamma$-ray bands. As a follow-up work of \citet{mao18}, this paper focuses on two issues. First, circular polarization of synchrotron radiation is not negligible. The circular polarization degree of a single electron as a function of $x$ is proportional to $1/\gamma$ when we fix the value of $x$, where $\gamma$ is the electron Lorentz factor, $x=\nu/\nu_c$, $\nu$ is the observational frequency, $\nu_c=\gamma^2\nu_L \rm{sin}(\alpha)$, $\nu_L=eB/m_ec$ is the gyroradius frequency, and $\alpha$ is the pitch angle. This indicates the possible detection of the circular polarization for some synchrotron sources in the optical and radio bands. Second, we should consider the synchrotron polarization radiative transfer in the optical and radio bands when synchrotron sources have dense media. \citet{jones77} studied the details of the synchrotron polarization radiative transfer. However, the electron energy distribution was assumed by a power law, and the thermal electron component was not considered. In this paper, we investigate the hybrid thermal-nonthermal electron contribution to the circular polarization of the synchrotron radiation and the radiative transfer processes of the synchrotron polarization.

It is expected that the polarization properties produced by the thermal-nonthermal electrons have wide applications to many kinds of synchrotron sources. Here, we propose some examples. \citet{covino17} detected the low-level linear polarization of AT 2017gfo/GW170817 during the kilonova phase. Although the measured polarization seems to be due to the Galactic dust along the line of sight, we think that future synchrotron polarization detections for electromagnetic counterparts of gravitational wave events can be applied to further investigate compact object merger systems and constrain detailed physics of gravitational wave events.
Fast radio burst (FRB) 140514 has the linear polarization limit of $<10\%$ and the circular polarization degree of $21\pm 7\%$ \citep{petroff15}. The development of the radio polarization measurements on FRBs are carried out to reveal the FRB origin. \citet{covino16} summarized the polarization observation of GRB prompt and afterglow emissions, and the GRB jet geometry related to the polarization was also mentioned.
\citet{steele17} presented the optical polarization results of some GRBs, and the linear polarization degrees are relatively low (see also the recent report of \citet{wiersema18} on GRB 180205A). Moreover, the circular polarization was detected in GRB 121024A \citep{wier14}.
Spectropolarimetric detection of either supernova Ia (SN Ia) or core-collapse supernova (CCSN) was performed \citep{tanaka12,porter16}.
The polarization properties can be helpful to further explore SN explosion mechanisms.
Thermal electrons cannot be ignored in the celestial objects mentioned above, and we should consider the thermal electron contribution to the synchrotron polarization. Moreover, some objects, such as kilonova, GRB, and SN as mentioned above, have dense environments \citep{delia07,kasen13,tanaka13,buckey18,yang18}. Polarization radiative transfer should be taken into account when we consider the polarization measurements in the optical and radio bands.
In this paper, we do not attempt to reproduce these different polarization properties with exact modeling parameters. However, these polarization detections of the different populations encourage us to further explore the possibility of applying our scenario, when we assume that all of these kinds of sources are dominated by the synchrotron radiation.

We illustrate the circular polarization results contributed by the relativistic thermal electrons in Section 2.1. The polarization radiative transfer is presented in Section 2.2. We discuss some possible applications in Section 3. The conclusions are given in Section 4.

\section{Radiation Transfer of Synchrotron Polarization by Hybrid Thermal-Nonthermal Relativistic Electrons}
Synchrotron polarization properties and radiation transfer were generally calculated by \citet{sazonov69} and \citet{jones77}. In particular, \citet{jones79} presented the spectral and polarization properties in the case that synchrotron radiation is fully produced by thermal electrons. In this paper, we focus on the polarization properties derived from the hybrid thermal/nonthermal electron distribution.
\subsection{Circular Polarization}
We follow the formula presentation and calculation given by \citet{sazonov69}, and the circular polarization of a single relativistic electron is
\begin{equation}
p_{c,s}=\frac{4\int^{\pi/2}_0\cos(\alpha)[xK_{1/3}(x)+\int^{\infty}_{x}K_{1/3}(z)dz]d\alpha}{3\int^{\pi/2}_0\gamma\sin(\alpha)[x\int^{\infty}_{x}K_{5/3}(z)dz]d\alpha},
\end{equation}
where $x=\nu/\nu_c$, $\nu$ is the observational frequency, $\nu_c=\gamma^2\nu_L \rm{sin}(\alpha)$, $\nu_L=eB/m_ec$ is the gyroradius frequency, $K(x)$ is the modified Bessel function, $\gamma$ is the electron Lorentz factor, and $\alpha$ is the pitch angle. The denominator in Equation (1) presents the total emission of an electron, and the numerator in Equation (1) presents the circularly polarized emission. The circular polarization degree can be explicitly calculated, and we plot the circular polarization degree as a function of $x$ in Figure 1. The degree steadily decreases when we increase $x$. This indicates that the circular polarization can be detected in low frequency and the detectors for the polarized signals in high-energy bands should be very sensitive. We also see that the circular polarization of a single electron is proportional to $1/\gamma$. The different results with different electron Lorentz factors $\gamma$ are illustrated. This indicates that the circular polarization produced from low-energy electrons has large degrees.

The total circular polarization degree by all relativistic electrons is given by
\begin{equation}
p_c=\frac{4\int^{\gamma_{max}}_{\gamma_{min}}\int^{\pi/2}_0\cos\alpha[(N_1+N_2)/\gamma][xK_{1/3}(x)+\int^{\infty}_{x}K_{1/3}(z)dz]d\alpha d\gamma}{3\int^{\gamma_{max}}_{\gamma_{min}}\int^{\pi/2}_0\gamma\sin\alpha(N_1+N_2)(x\int^{\infty}_{x}K_{5/3}(z)dz)d\alpha d\gamma},
\end{equation}
where we assume $\gamma_{min}=1$ and $\gamma_{max}=10^6$, where $N_1$ and $N_2$ are the thermal and nonthermal energy distributions of the relativistic electrons, respectively. Thus, the total circular polarization contains the contribution from both thermal and nonthermal electrons. This also allows us to investigate the polarization differences between thermal and nonthermal electrons in many physical cases with wide applications. We utilize the hybrid electron energy distribution proposed by \citet{giannios09}. The electron energy distributions are
\begin{equation}
 N_{1}(\gamma)=C_e\gamma^2\rm{exp}(-\gamma/\Theta)/2\Theta^3~~~for~~~\gamma \le \gamma_{th},
\end{equation}
and
\begin{equation}
N_{2}(\gamma)=C_e\gamma_{th}^2\rm{exp}(-\gamma_{th}/\Theta)(\gamma/\gamma_{th})^{-n}/2\Theta^3~~~for~~~\gamma> \gamma_{th},
\end{equation}
where $C_e$ is a constant, $n$ is the power-law index of the electron energy distribution, and $\gamma_{th}$ is the conjunctive electron Lorentz factor that has the range of $\gamma_{min}<\gamma_{th}<\gamma_{max}$. We take $n=2.5$ during our calculations. The electron characteristic temperature is defined by $\Theta=kT_e/m_ec^2$, and $T_e$ is the thermal temperature of the relativistic electrons.
When $\gamma_{th}$ is approaching to $\gamma_{min}$, the circular polarization can be dominated by the nonthermal electrons. While the thermal electrons dominate the circular polarization if $\gamma_{th}$ is approaching to $\gamma_{max}$. We have investigated the properties of the electron energy distributions as a function of $\gamma_{th}$ and $\Theta$ in the Figure 1 of \citet{mao18}. We note that the electron drift velocity is not included in our calculation. If the drift velocity is relativistic, an additional Lorentz factor $\Gamma_d$ can be adopted in the calculation as an average electron drift effect.

We show the synchrotron circular polarization degrees produced by the electrons with the hybrid energy distribution as a function of $\gamma_{th}$ in Figure 2.
We suggest some cases in the radio (5 GHz), optical (5500\AA), X-ray (5 keV), and $\gamma$-ray (5 MeV) bands, and we obtain the different circular polarization degrees in these observational energy bands. Electron temperature as an important parameter in the hybrid electron energy distribution is also considered in the polarization calculations. The circular polarization degree is varied with the conjunctive Lorentz factor. We note the following properties in our results: (1) One common feature is that the strong transition of the polarization degree is shown in most cases in Figure 2. This kind of transition is due to the term of $exp(-\gamma_{th}/\Theta)$ given in the electron energy distribution Equations (3) and (4). If we choose a high electron temperature, the circular polarization degree has a smooth transition when we change $\gamma_{th}$. On the other hand, if the electron temperature is relatively low, the circular polarization degree can have a very sharp transition when we change $\gamma_{th}$. This property is similar to that of the linear polarization presented by \citet{mao18}. (2) It is shown in Figure 2 that the circular polarization degrees are dominated by the thermal, nonthermal, and thermal-nonthermal transition phases, respectively. We rewrite Equation 2 as $p_c=(V_{thermal}+V_{nonthermal})/(I_{thermal}+I_{nonthermal})$. When the thermal radiation dominates one case, we have $V_{thermal}\gg V_{nonthermal}$ and $I_{thermal}\gg I_{nonthermal}$. Thus, $p_c=V_{thermal}/I_{thermal}$. However, in this case, it may happen that $(V_{thermal}/I_{thermal})<(V_{nonthermal}/I_{nonthermal})$.
Although the detected radiation and polarization can be dominated by the thermal electrons, we cannot distinguish the polarized emission dominated by the thermal or nonthermal electrons. The same situation occurs in the nonthermal radiation dominated case. (3) Although the circular polarization degree of a single electron is proportional to $\gamma$ as we see in Figure 1, we present $x=\nu/\nu_c$, where $\nu_c$ is proportional to $\gamma^2$. Therefore, it is not straightforward to qualitatively analyze the circular polarization degree of the total relativistic electrons presented by Equation (2). Quantitative computation of Equation (2) in different observational energy bands is necessary. The different polarization degrees with different electron temperatures in each energy band are specified in Figure 2. (4) It seems that the circular polarization can be detectable in the low-energy bands in the case, in which the plasma temperature is low. We should consider the linear polarization results of \citet{mao18} and perform the polarization radiative transfer processes to obtain linear and circular polarization, and compare our modeling results with observational measurements.

\subsection{Polarization Radiative Transfer}
We follow the calculation of the radiation transfer coefficients given by \citet{sazonov69} and \citet{jones77}. The absorption coefficient of the total radiation is
\begin{equation}
k_I=-\frac{\sqrt{3}e^2}{4\pi mc\nu^2}\int^{\gamma_{max}}_{\gamma_{min}}\int^{\pi/2}_{0} \gamma^2\nu_L\sin\alpha [x\int^{\infty}_xK_{5/3}(t)dt] \frac{\partial}{\partial \gamma}[\frac{N_e(\gamma)}{\gamma^2}] d\gamma d\alpha,
\end{equation}
where $N_e(\gamma)=N_1(\gamma)+N_2(\gamma)$. This means that the synchrotron absorption is contributed by both thermal and nonthermal electrons. In this paper, we focus on the polarization radiative transfer. However, if we consider the synchrotron self-absorption process, the effect from the thermal electrons also seems important \citep{warren18}.
The absorption coefficient of the linear polarized component is
\begin{equation}
k_Q=-\frac{\sqrt{3}e^2}{4\pi mc\nu^2}\int^{\gamma_{max}}_{\gamma_{min}}\int^{\pi/2}_{0} \gamma^2\nu_L\sin\alpha [xK_{2/3}(x)]\frac{\partial}{\partial \gamma}[\frac{N_e(\gamma)}{\gamma^2}] d\gamma d\alpha,
\end{equation}
and the absorption coefficient of the circular polarized component is
\begin{equation}
k_V=-\frac{e^2}{\sqrt{3}\pi mc\nu^2}\int^{\gamma_{max}}_{\gamma_{min}}\int^{\pi/2}_{0} \gamma\nu_L\cos\alpha [xK_{1/3}(x)+\int^{\infty}_xK_{1/3}(z)dz]\frac{\partial}{\partial \gamma}[\frac{N_e(\gamma)}{\gamma^2}] d\gamma d\alpha.
\end{equation}
The coefficients of rotativity and the convertibility are
\begin{equation}
k^*_V=-\frac{2e^2}{mc\nu^2}\int^{\gamma_{max}}_{\gamma_{min}}\int^{\pi/2}_{0} \nu_L\cos\alpha (\gamma \rm{ln}\gamma) \frac{\partial}{\partial \gamma}[\frac{N_e(\gamma)}{\gamma^2}] d\gamma d\alpha,
\end{equation}
and
\begin{equation}
k^*_Q=-\frac{3^{1/3}e^2}{2^{4/3}mc\nu^2}\int^{\gamma_{max}}_{\gamma_{min}}\int^{\pi/2}_{0} x\nu_L\sin\alpha \{\int^\infty_0 z\cos[z(3\nu/2\nu_c)^{2/3}+z^3/3]dz\}\gamma^2\frac{\partial}{\partial \gamma}[\frac{N_e(\gamma)}{\gamma^2}] d\gamma d\alpha,
\end{equation}
respectively.
We can write the radiation transfer matrix as
\begin{eqnarray}
\left(
\begin{array}{cccc}
d/dl+k_I & k_Q      & 0        & k_V \\
k_Q      & d/dl+k_I & k^\ast_V & 0 \\
0        &-k^\ast_V & d/dl+k_I & k^\ast_Q \\
k_V      & 0        & -k^\ast_Q& d/dl+k_I
\end{array}     \right)
\left(
\begin{array}{c}
I  \\
Q  \\
0  \\
V
\end{array}     \right)
= \left(
\begin{array}{c}
0  \\
0  \\
0  \\
0
\end{array}     \right).
\end{eqnarray}
Here, we assume that the emission coefficients are zero. We traditionally treat the polarization as though it has a position angle of zero degrees so that we simply put $U=0$. The optical depth element can be defined as $d\tau=k_I dl$. We can solve Equation (10) and obtain the linear and circular polarization degrees as a function of the optical depth.

We take the intrinsic linear polarization results derived by the hybrid electron energy distribution from \citet{mao18}. We combine the linear polarization results of \citet{mao18} with the circular polarization results from this paper. Then, we perform the radiative transfer processes mentioned above and choose some cases of the linear and circular polarization degrees as a function of the optical depth in the optical (5500\AA) and radio (5 GHz) bands. The final results are shown in Figure 3, and the details are written in the caption. We simply ignore the results of the circular polarization degree less than $10^{-4}$, because low-level circular polarization cannot be detected by radio and optical telescopes in our research field. We propose some examples to apply our results to different celestial objects in the following section.

\section{Possible Applications}

We present some general polarization properties in the synchrotron radiation framework in this paper. Aiming for one accurate prediction or an explanation of one certain observational polarization result with some special modeling parameters is not our goal. However, the synchrotron polarization produced by the hybrid electron energy distribution has wide applications. We attempt to discuss some possible applications. In the following preliminary results that we obtain, we do not consider any extrinsic effects, such as magnetic field morphology, jet structure with off-axis, and depolarization, superposed to the intrinsic polarization degrees.

The electromagnetic counterpart of the gravitational wave event AT 2017gfo/GW 170817 was confirmed as a binary neutron star merger \citep{abbott17}. One attempt of linear polarization detection during the kilonova phase was presented by \citet{covino17}. The latest findings explore the possibility that GW 170817 may have originated from a GRB inside a cocoon structure \citep{mooley18} or from structured relativistic jets \citep{lyman18}. One of our results presented by the solid line in the left panels of Figure 3 corresponds to a linear polarization in the case of $T_e=10^{13}$ K and $B=1$ G. We may apply this case to investigate the linear polarization of AT 2017gfo/GW 170817, if the synchrotron radiation can be applied to this source during the GRB afterglow phase. In the meantime, the case of $T_e=10^{12}$ K and $B=1$ G denoted by the dashed line in the left panels of Figure 3 is also possible for this study. \citet{gill18} have discussed the polarization properties from jet/cocoon models. In our paper, the GRB afterglow polarization linked to AT 2017gfo/ GW 170817 may suffer a strong absorption due to the surrounding cocoon, while structured jets may have less absorption effects on the polarization. We also predict a relatively high degree of linear polarization in the radio band that corresponds to the case of $T_e=10^{12}$ K and $B=1$ G as shown by the dashed line in the right panels of Figure 3, and the radio polarization measurements are strongly suggested when the afterglow is bright enough.

FRBs have been detected in the radio band. Circular polarization of FRB 140514 was measured at a level of $(21\pm7)\%$ degree, while linear polarization is less than 10\% as a 1$\sigma$ upper limit \citep{petroff15}. We apply our model to self-consistently get both circular and linear polarization degrees of FRB 140514. The case of $B=10^{10}$ G, $T_e=10^8$ K, and $\gamma_{\rm{th}}=10^4$ shown as the solid line in the right panels of Figure 3 seems to be one solution to explain the polarization observations. Thus, we may consider FRB 140514 to be associated with strongly magnetized sources. A neutron star can be one candidate for FRB origin. Here, we assume that the polarized emission is pointing toward observers, as usual neutron star models assumed, but we do not include additional jet view-angle effects.
We note one recent theoretical work on the polarization detection of neutron stars in the soft X-ray band \citep{yatabe17}, and the relatively low electron temperature from our analysis indicates that the radiation of FRB 140514 may not occur near the neutron star magneto-atmosphere above the neutron star surface.
On the other hand, the alternative case of $B=1$ G, $T_e=10^{10}$ K, and $\gamma_{th}=3\times 10^3$ in our model can obtain the circular polarization degree of 20\%. However, in this case, the intrinsic linear polarization degree is about 76\%. This seems to be in contrast with the observation of FRB 140514. Therefore, we are inclined to believe from our polarization analysis that FRB may have the neutron star origin.
Besides FRB 140514, FRB 110523 has the linear polarization degree of $(44\pm 3)\%$, but the the circular polarization degree of about 23\% due to instrumental leakage was not confirmed to be an astronomical result \citep{masui15}.
FRB 081507 has the high linear polarization degree of 80\% \citep{ravi16} and FRB 150215 has the linear polarization degree of $(43\pm 5)\%$ \citep{petroff17}.
If we can detect both linear and circular polarization for FRBs, we may further constrain the FRB physics.

The typical linear polarization degree detected in GRB optical afterglows is about a few percent with some noticeable exceptions \citep{covino16}. A linear polarization degree of 4\% and a circular polarization degree of 0.6\% were detected in the optical band in GRB 121024A \citep{wier14}. If we consider one case of $B=1$ G and $T_e=10^{12}$ K in our model, which is presented by the dashed line in the left panels of Figure 3, we may roughly get the same order of circular polarization degree. When we consider the low-level linear polarization from observations, we may further consider several additional possibilities. First, a lower linear polarization degree can be obtained if the optical depth is sightly larger than 1. This is reasonable since long GRBs are assumed to have dense environments. Second, the GRB jet off-axis effect and the magnetic field morphology can decrease the linear and circular polarization degree \citep{granot03,nava16}. Third, depolarization effects can also induce a lower linear polarization.

Supernova explosion releases shocks. In the early supernova remnant (SNR) phase, nonrelativistic shocks sweep up the surrounding medium. Shocks not only accelerate particles but also amplify the magnetic field \citep{duffell17,kundu17,xu17,sato17}. Synchrotron polarization maps of SNRs have been investigated \citep{band16,petruck17}. However, the synchrotron polarization contributed by the thermal electrons is not considered. In this paper, we compute radio polarization for SNRs in the case of $B=10^{-3}$ G, $T_e=10^{10}$ K, and $\gamma_{\rm{th}}=10^4$. In the meantime,
we also compute optical polarizations for SNRs in the case of $B=10^{-3}$ G, $T_e=10^{12}$ K, and $\gamma_{\rm{th}}=10^4$. Although SNR polarization properties have been measured from optical spectropolarimetry \citep{tanaka12,porter16} and our results do not concern any absorption/emission lines, we expect that our polarimetric results can also be helpful to explore SN explosion mechanisms.

We note that magnetic fields can also be effectively amplified by relativistic shocks in blazar jets \citep{mizuno14,aroudo15,kino17}. In our paper, the case of $B=10^{-3}$ G, $T_e=10^{12}$ K, and $\gamma_{\rm{th}}=10^4$ in the optical band may also have possible applications for the research of the blazar optical polarization. Besides the absorption from the radiation transfer, the turbulent depolarization on different optical bands can have effects on the linear polarization degrees \citep{guo17}. The work in this paper may also be useful to explain the recent polarization monitoring results at millimeter bands \citep{agudo18,thum18}.

We may also apply our model to other celestial populations. For example, we have strong evidence for the coexistence of thermal and nonthermal electrons in accretion flow \citep{yuan14}. The polarization measurements were performed in the radio and infrared bands during the flaring phase of the Galactic center $\rm{Sgr~A}^*$ \citep{sha15,liu16}. In particular, the circular polarization has been detected at submillimeter bands \citep{munoz12}. Thermal electrons and radiative transfer have been considered in this research topic \citep{yuan03,li17b}.
We note that the hybrid thermal/nonthermal electron energy distribution and the strong absorption of polarized photons in our model should be important when we suggest to perform Event Horizon Telescope (EHT) polarization observations. The circular polarization of pulsar wind nebula (PWN) can also be further checked if we consider the thermal electron contribution \citep{bu18}. We may even consider performing our model for the diffuse synchrotron emission if the electrons in the interstellar medium also have a thermal component \citep{herron18}.

Although we develop our model for some types of celestial objects with polarized emission in this paper, it is still difficult to describe some special polarization properties in observations. For instance, FRB 121102 has almost 100\% linear polarization with strongly varied Faraday rotation \citep{michilli18}. It is hard to reproduce the observed 100\% linear polarization degree by using the intrinsic synchrotron polarization when we assume that FRB 121102 is related to a neutron star. We speculate some additional issues for this special event. First, if the electrons of FRB 121102 have an extremely large number of Lorentz factors and the pitch angle is extreme along the radiation opening angle, it seems that the emission is along the direction of the magnetic field. Thus, we obtain an extremely large linear polarization and the circular polarization is zero. It may happen in the case that the neutron star has very fast spin. Second, the plasma lensing effect in FRB host galaxies is proposed because the amplitudes of FRB emission can be violently modulated \citep{cordes17}. In the case that the amplitudes are extremely anisotropic, the detected linear polarization may almost reach a degree of 100\%. Third, besides the simple rotating diploe model of neutron star, the open magnetic field lines near the light cylinder can be twisted. This is one example of complicated magnetic field configuration that can be modify the linear polarization. Fourth, we expect more observational samples with different possible scenarios to systematically analyze FRB physics \citep{vieyro17,caleb18}. On the other hand, we note that synchrotron radiation is not a unique radiation mechanism to describe the polarization properties. Jitter radiation, the relativistic electron radiation in random and small-scale magnetic fields, can also be a possible radiation mechanism when we consider more complicated observational phenomena. The high-degree linear polarization can be achieved, in general, by the jitter polarization, and the highly
varied Faraday rotation indicates the existence of strong turbulence \citep{mao13,mao17}.

\section{Conclusions}

The thermal signature in the radiative spectrum from the contribution of the thermal electrons has been presented \citep{jones79,stawarz08,giannios09}. From an observational point of view, the identification of the spectral thermal feature implies the existence of the thermal electrons. Polarization measurements provide further constrains on the nature of observational synchrotron sources. It is more important that our model provides a possibility to explain the diversity of the observational polarization degrees in the same synchrotron population. When we consider the polarization in high-energy bands as mentioned in \citet{mao18}, multiwavelength polarization observations are encouraged in the future.
We need synthetical analysis from light curves, spectra, and polarizations to confirm the detailed physics of synchrotron sources.

\acknowledgments
We are grateful to the referee for a careful review and helpful suggestions.
We thank F. Yuan, P. D'Avanzo, and Z.-Q. Shen for their useful discussion.
J.M. is supported by the National Natural Science Foundation of China 11673062, the Hundred Talent Program of Chinese Academy of Sciences, the Major Program of Chinese Academy of Sciences (KJZD-EW-M06), and the Oversea Talent Program of Yunnan Province. S.C. acknowledges partial funding from ASI-INAF grant I/004/11/3.
J.W. is supported by the National Natural Science Foundation of China (11573060 and 11661161010).

\clearpage




\clearpage

\begin{figure}
\includegraphics[scale=0.6]{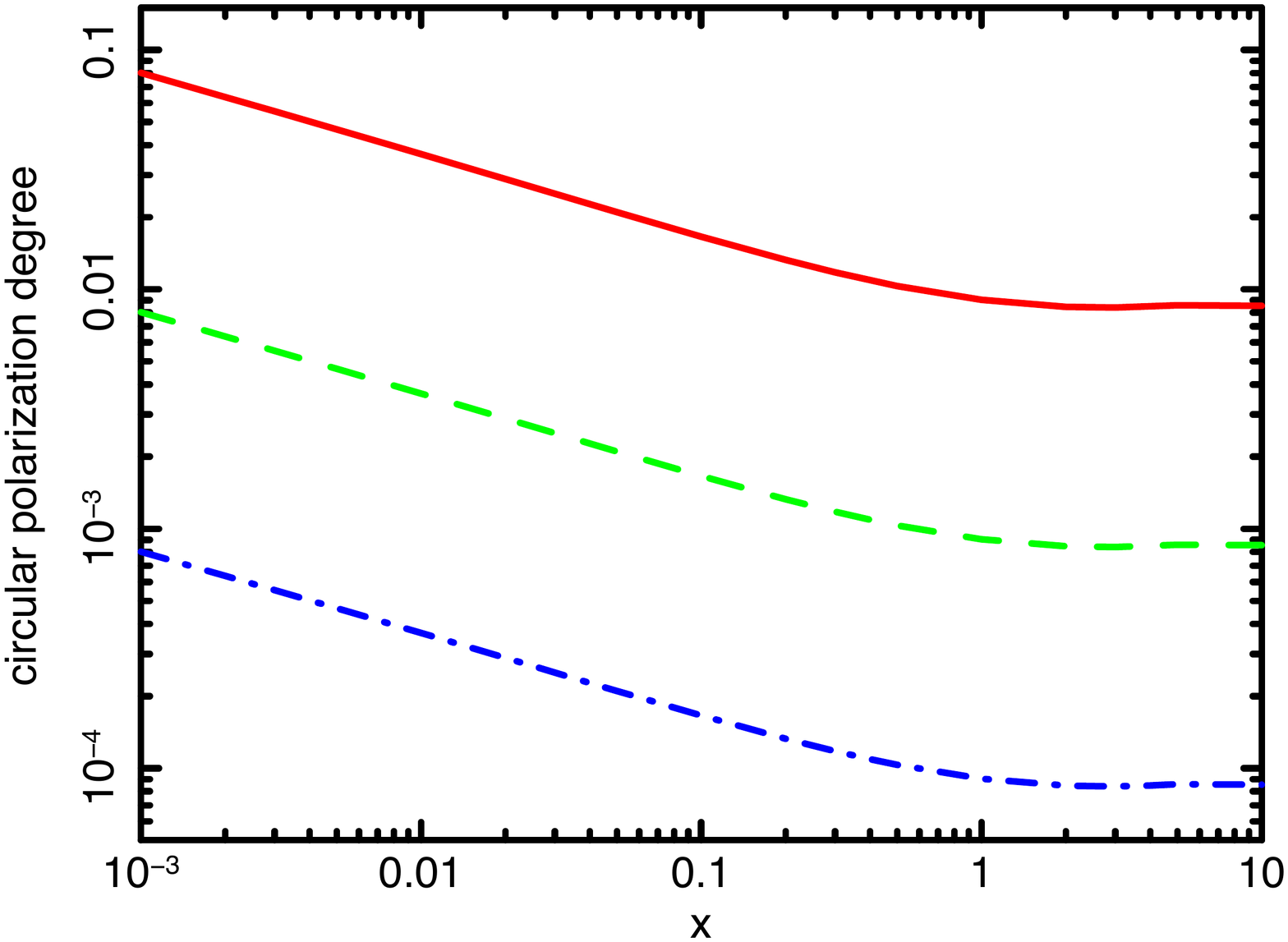}
\caption{Circular polarization degree of a single electron electron. Here, we directly take a number of $x$ in the calculation with a range from $10^{-3}$ to 10. The results are presented by the solid, dashed, and dotted-dashed lines when we take $\gamma=10^2,~10^3,~\rm{and}~10^4$, respectively.
\label{fig1}}
\end{figure}

\begin{figure}
\includegraphics[scale=0.3]{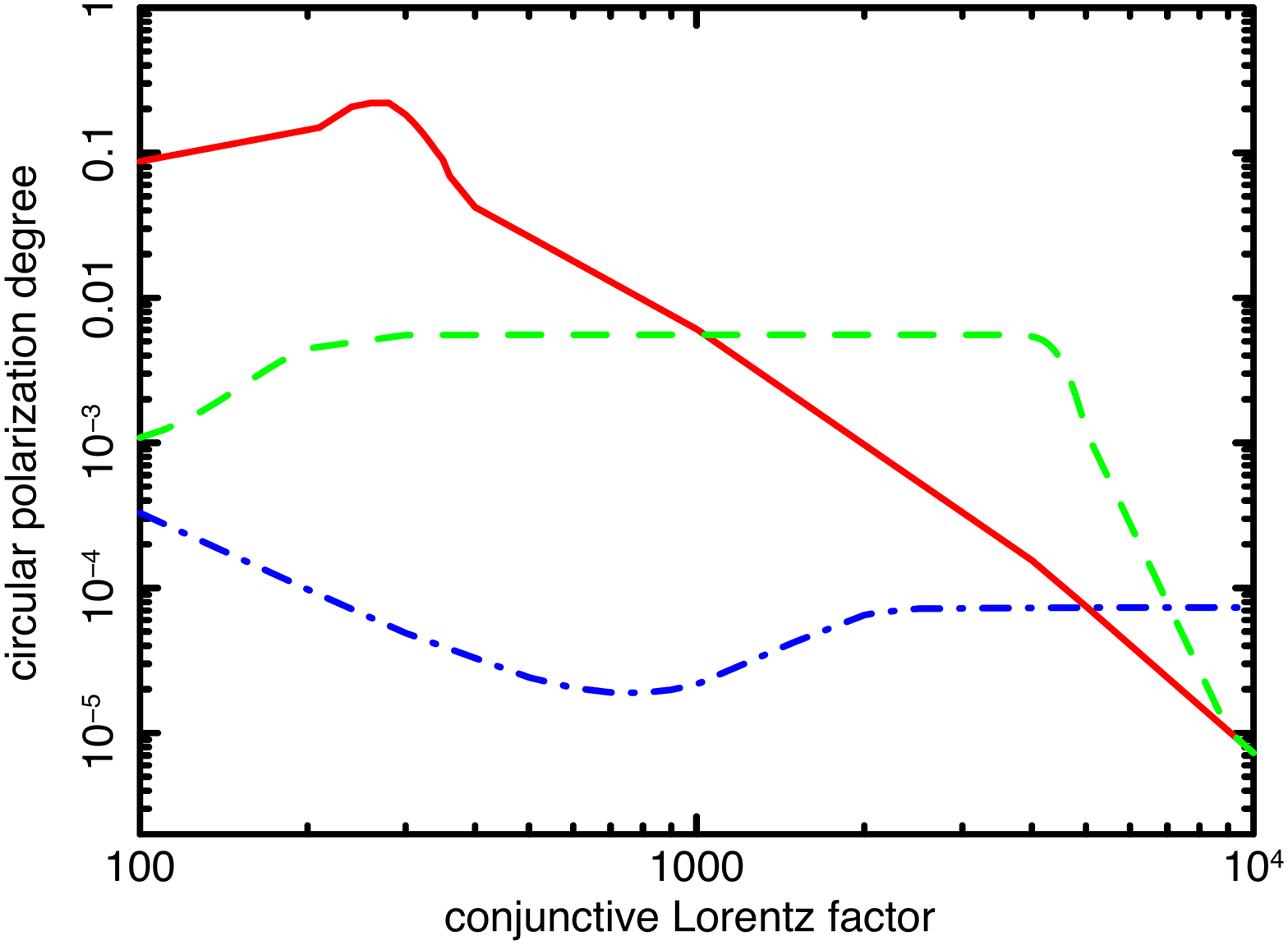}
\includegraphics[scale=0.3]{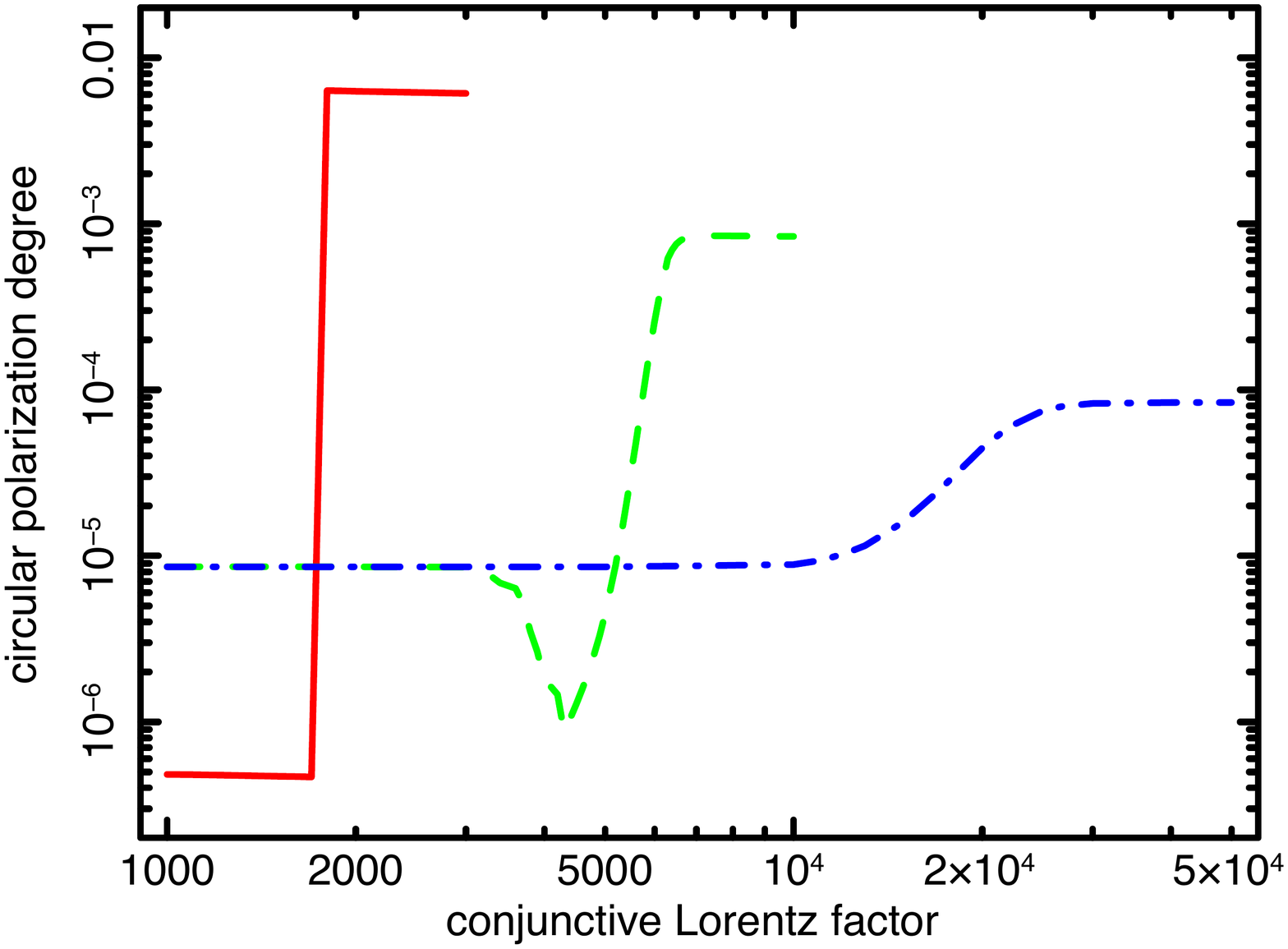}
\includegraphics[scale=0.3]{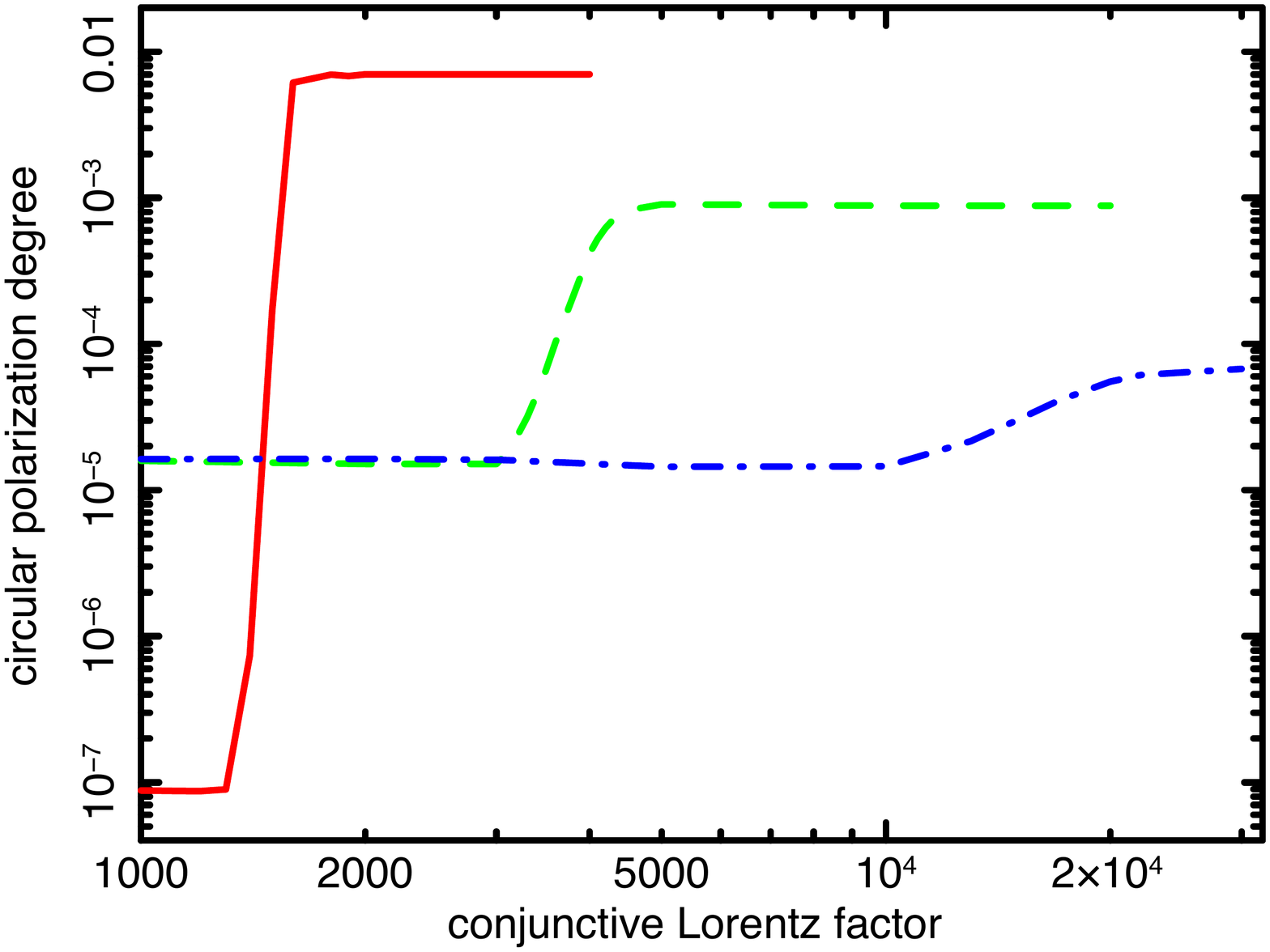}
\includegraphics[scale=0.3]{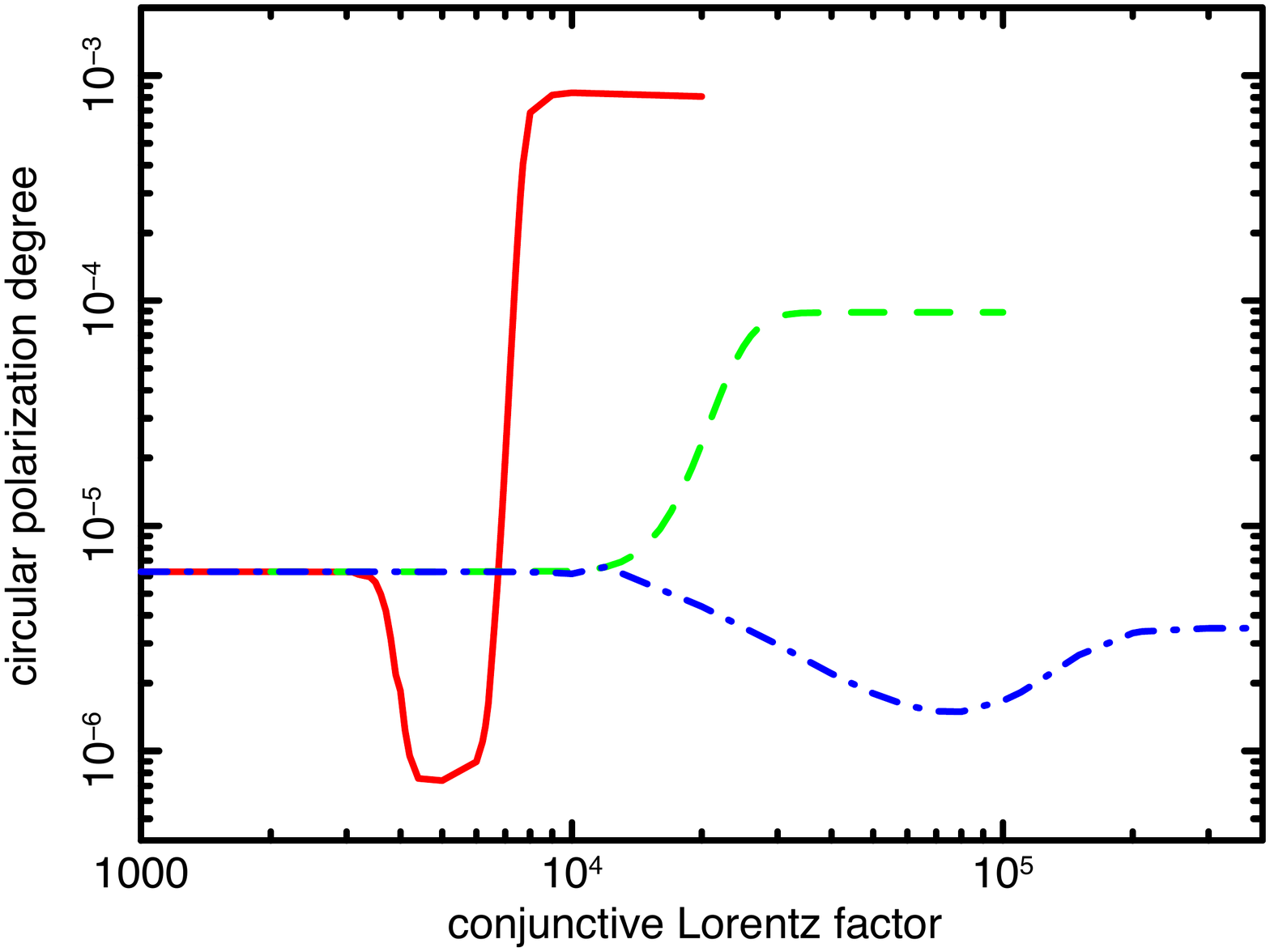}
\caption{Intrinsic circular polarization degree as a function of the electron conjunctive Lorentz factor $\gamma_{th}$. Top left panel: the circular polarization degree in the radio band (5 GHz) with $B=1$ G. The results are presented by the solid, dashed, and dotted-dashed lines when we take $T_e=10^{10},~10^{11},~\rm{and}~10^{12}$ K, respectively. Top right panel: the circular polarization degree in the optical band (5500\AA) with $B=1$ G. The results are presented by the solid, dashed, and dotted-dashed lines when we take $T_e=10^{11},~10^{12},~\rm{and}~10^{13}$ K, respectively. Bottom left panel: the circular polarization degree in the X-ray band (5 keV) with $B=10^4$ G. The results are presented by the solid, dashed, and dotted-dashed lines when we take $T_e=10^{11},~10^{12},~\rm{and}~10^{13}$ K, respectively. Bottom right panel: the polarization degree in the gamma-ray band (5 MeV) with $B=10^6$ G. The results are presented by the solid, dashed, and dotted-dashed lines when we take $T_e=10^{12},~10^{13},~\rm{and}~10^{14}$ K, respectively.
\label{fig2}}
\end{figure}

\begin{figure}
\includegraphics[scale=0.3]{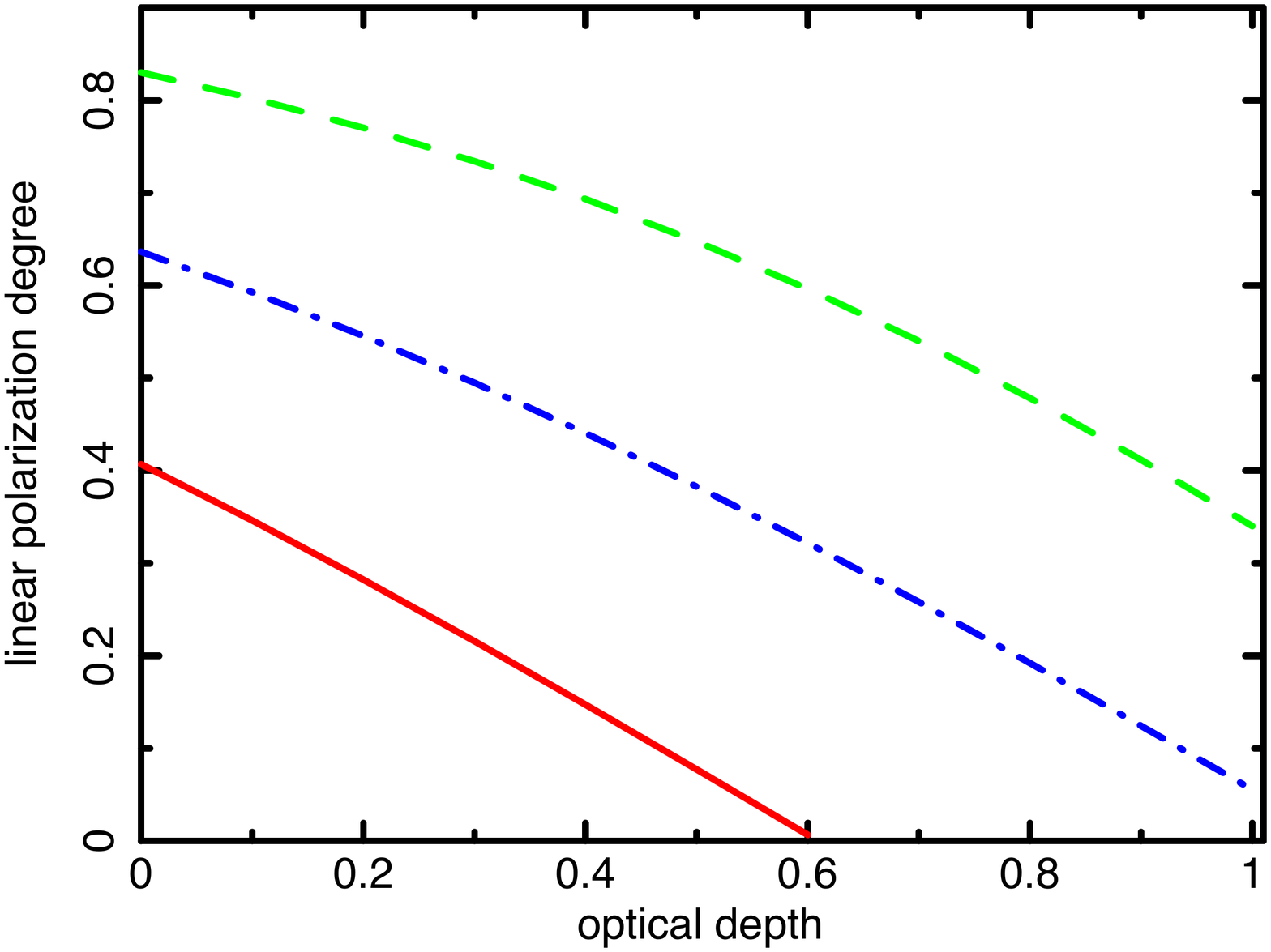}
\includegraphics[scale=0.3]{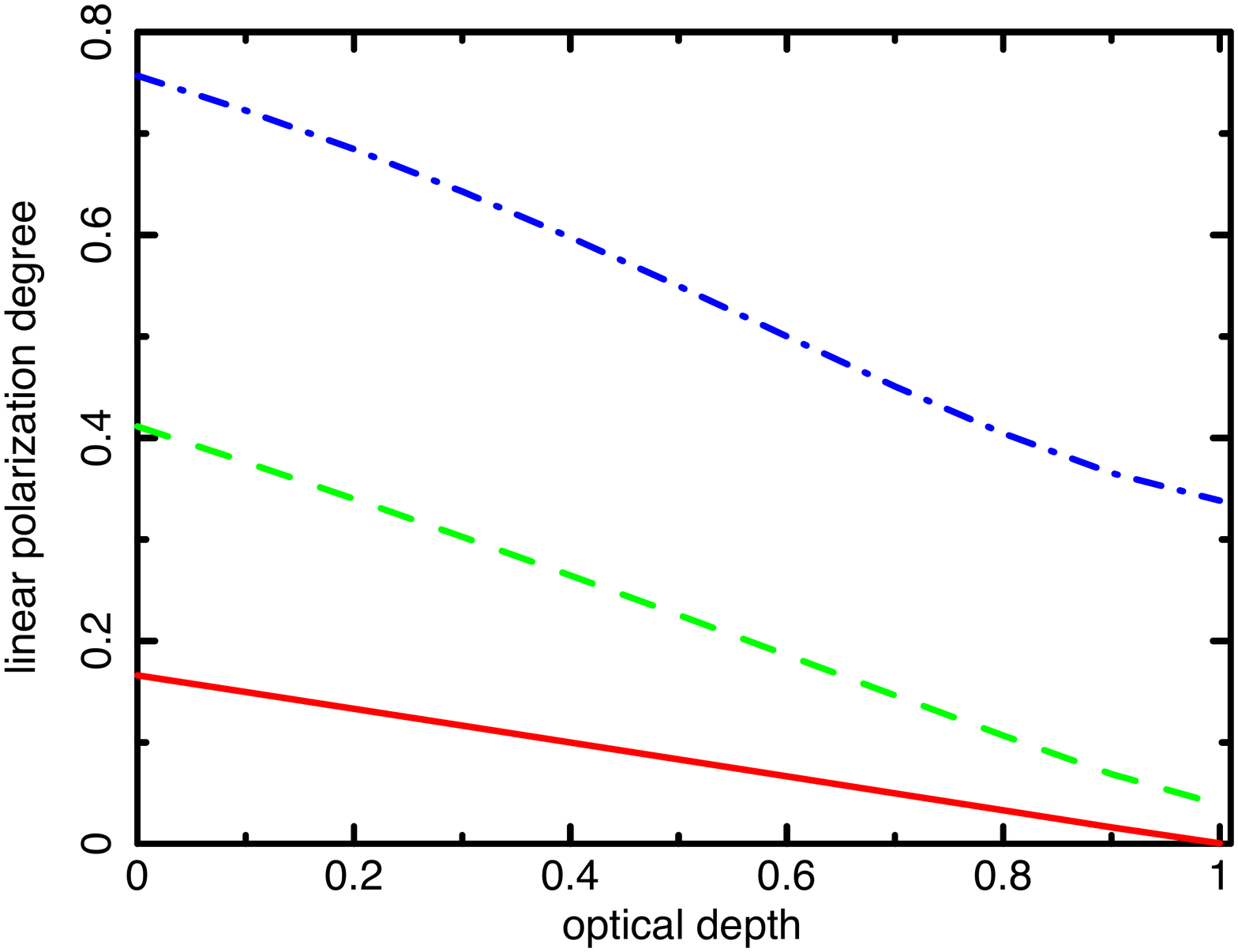}
\includegraphics[scale=0.3]{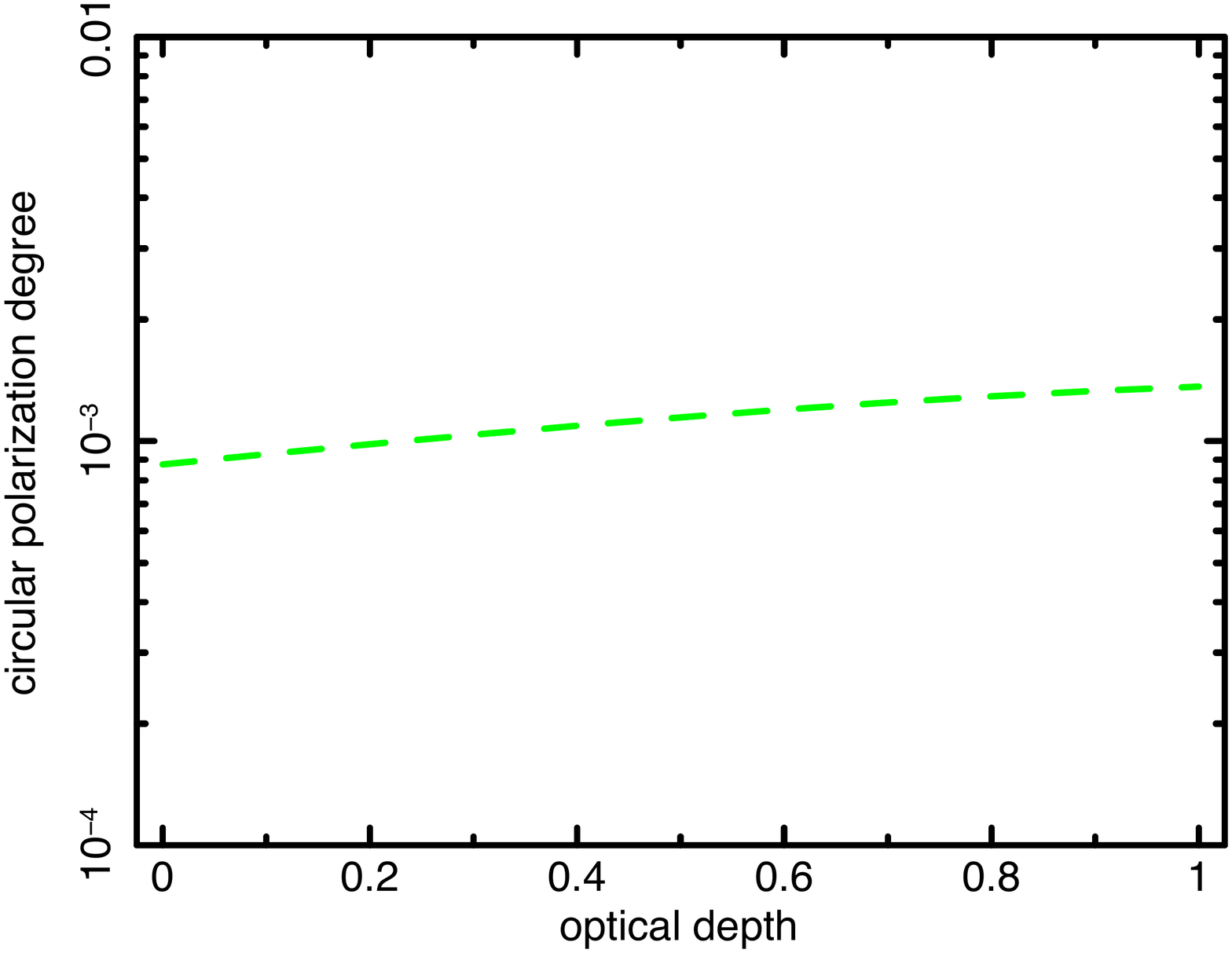}
\includegraphics[scale=0.3]{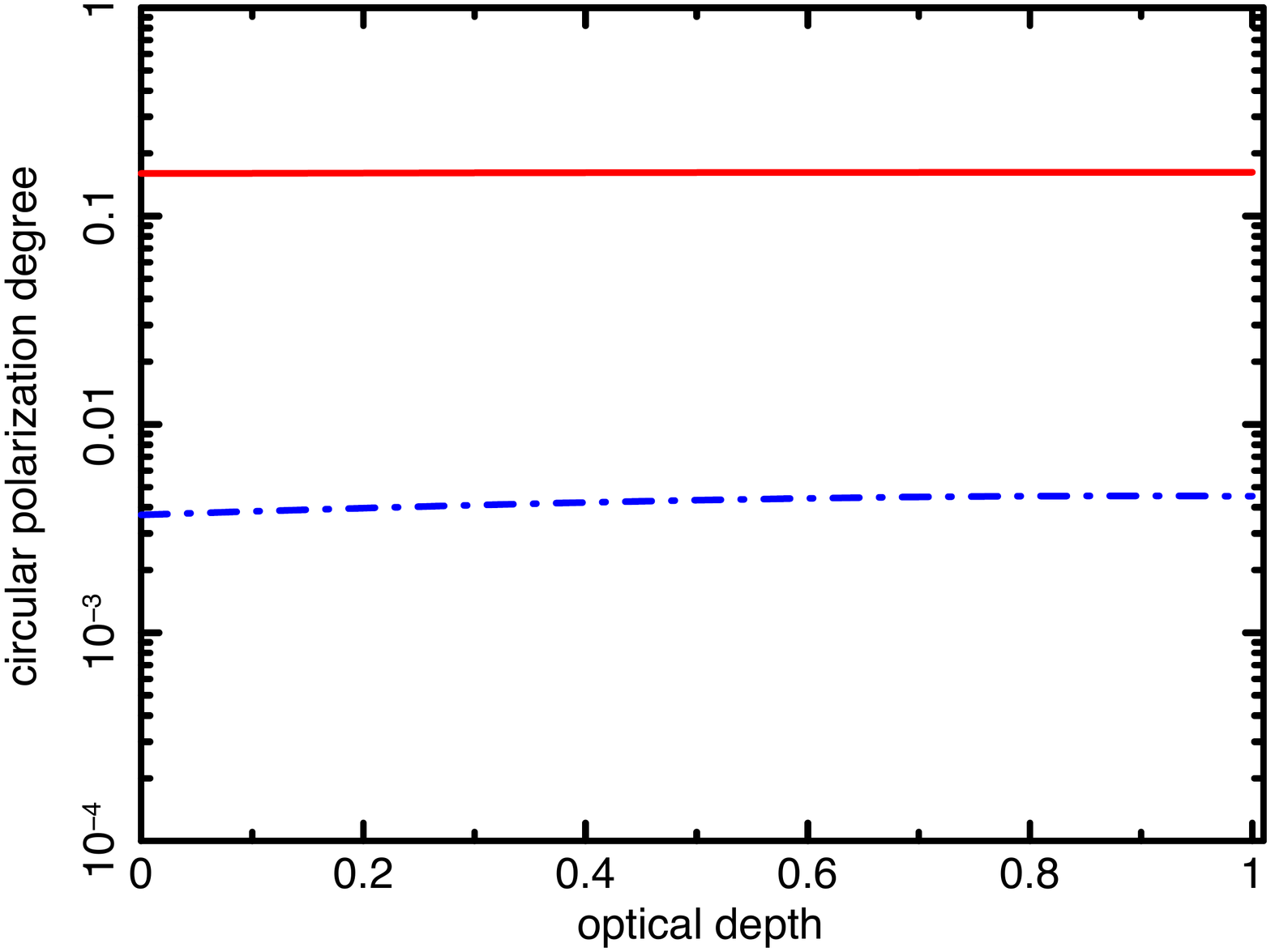}
\caption{Linear and circular polarization degrees as a function of optical depth. The electron conjunctive Lorentz factor is adopted as $\gamma_{\rm{th}}=10^4$. Top left panel: the linear polarization degree in the optical band (5500\AA). The results are presented by the solid ($T_e=10^{13}$ K and $B=1$ G), dashed ($T_e=10^{12}$ K and $B=1$ G), and dotted-dashed ($T_e=10^{12}$ K and $B=10^{-3}$ G) lines, respectively. Top right panel: the linear polarization degree in the radio band (5 GHz). The results are presented by the solid ($T_e=10^8$ K and $B=10^{10}$ G), dashed ($T_e=10^{12}$ K and $B=1$ G), and dotted-dashed ($T_e=10^{10}$ K and $B=10^{-3}$ G) lines, respectively.
Bottom left panel: the circular polarization degree in the optical band (5500\AA). The results are presented by the solid ($T_e=10^{13}$ K and $B=1$ G), dashed ($T_e=10^{12}$ K and $B=1$ G), and dotted-dashed ($T_e=10^{12}$ K and $B=10^{-3}$ G) lines, respectively.
Bottom right panel: the circular polarization degree in the radio band (5 GHz). The results are presented by the solid ($T_e=10^8$ K and $B=10^{10}$ G), dashed ($T_e=10^{12}$ K and $B=1$ G), and dotted-dashed ($T_e=10^{10}$ K and $B=10^{-3}$ G) lines, respectively.
In the above plots, we neglect the cases with the circular polarization degree less than $10^{-4}$.
\label{fig3}}
\end{figure}

\end{document}